\documentstyle[aps,prb,twocolumn,graphicx]{revtex}

\begin{document}

\twocolumn[\hsize\textwidth\columnwidth\hsize\csname
@twocolumnfalse\endcsname

\title{Longitudinal conductivity and transverse charge
redistribution in coupled quantum wells subject to in-plane magnetic
fields}

\author{J.\ Koloren\v c, L.\ Smr\v cka, and P.\ St\v reda}

\address{Institute of Physics ASCR, Cukrovarnick\'a 10, 162 53 Praha
6, Czech Republic}
\draft

\date{\today}
\maketitle

\begin{abstract}
In double quantum wells electrons experience a Lorentz force oriented
perpendicular to the structure plane when an electric current is
driven perpendicular to the direction of an in-plane magnetic field.
Consequently, the excess charge is accumulated in one of the wells. The
polarization of a bilayer electron system and the corresponding Hall
voltage are shown to contribute substantially to the in-plane
conductivity.
\end{abstract}


\vskip2pc]
\section{Introduction}
A finite size, in the growth direction, of bilayer electron systems
confined in double quantum wells leads to strong orbital effects of an
applied in-plane magnetic field. As a result, an oscillation appears
in the field dependent density of states, corresponding to the
depopulation of the antibonding subband at a critical field $B_{\|} =
B_{c,1}$, and to the splitting of the Fermi see into two separated
electron sheets at a second critical field $B_{c,2}$. The variation of
the density of states with $B_{\|}$ is reflected in magnetoresistance
traces recorded as functions of $B_{\|}$. Critical fields $B_{c,1}$
and $B_{c,2}$ can, therefore, be determined experimentally
\cite{si,ku,ju,mak}.

The connection between the magnetoresistance $\rho(B_{\|})$ and the
density of states $g(B_{\|})$ employed in this papers is based on an
assumption that the scattering rate is proportional to the density of
states.  Unlike $g(B_{\|})$, which is determined exclusively by the
electron energy spectra, the $\rho(B_{\|})$ depends also on the
electron scattering mechanism and on the mutual orientation of the
electric current and the magnetic field. Therefore, the
proportionality between $\rho(B_{\|})$ and $g(B_{\|})$ is only
approximate.

A conventional approach describing transport in double quantum wells
subject to in-plane magnetic fields relies on one-electron
approximation and semiclassical Boltzmann equation \cite{Raichev}.
This approach is justified in the case of ``parallel'' conductivity
$\sigma_{\|}$ related to magnetoresistance by
$\rho_{\|}=1/\sigma_{\|}$, i.e., when an electric current is driven
along the magnetic field $\left(\vec{j}\; \|\, \vec{B}\right)$ and no
Lorentz force acts on electrons.

Here we present a theoretical scheme which includes the
electron-electron interactions and allows to evaluate the
``transverse'' conductivity $\sigma_{\perp}$ $\left(\rho_{\perp}=
1/\sigma_{\perp}\right)$ corresponding to an in-plane electric
current flowing perpendicular to the field direction, $\vec{j}\, \perp
\vec{B}$.  In this case, electrons are driven by a Lorentz force
perpendicular to the 2D-planes and the electric charge is transferred
from one well to the other in the current carrying state.  The Hall
contribution to $\rho_{\perp}$, obtained only in our approach, is
comparable in magnitude to the original $\rho_{\perp}$, and hence is
important for the analysis of the transport data.
\section{Electronic structure  and  transport coefficients}
\begin{figure}[hbt]
\includegraphics[width=7.5cm]{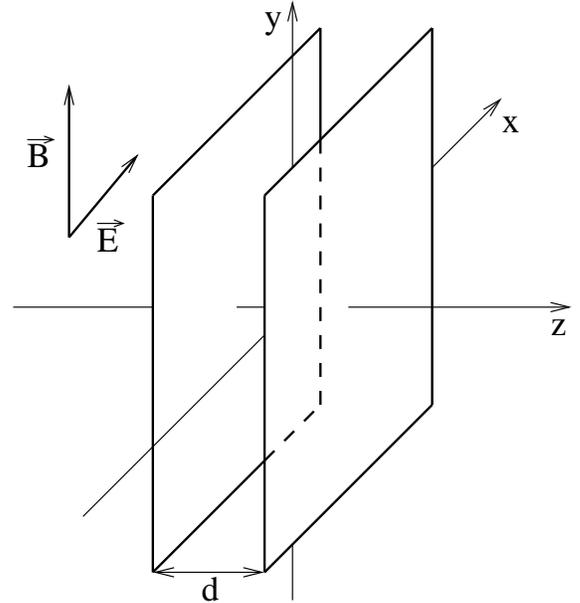}
\caption{Schematic picture of a double-layer system.  Directions of
applied electric and magnetic fields are indicated.}
\label{fig1}
\end{figure}

We consider two coupled, strictly two-dimensional (2D) electron layers
confined in very narrow potential wells at the distance $d$ (see
Fig.~\ref{fig1}).  The vector potential, $\vec{A}=(zB,0,0)$, is used to
describe the influence of the in-plane magnetic field,
$\vec{B}=(0,B,0)$, on the electronic structure. 
Only the lowest bound
states of the left and right wells are used to construct the model
Hamiltonian $\hat{H}_0$ of the system \cite{Hu}.  The Hamiltonian is
diagonal in the wavevector $\vec{k}\equiv (k_x,k_y )$ and takes a
matrix form
\begin{equation}
\langle\vec{k}|\hat{H}_0|\vec{k}'\rangle=\delta_{\vec{k},\vec{k}'}
\pmatrix{
\epsilon_L(\vec{k}) &     t         \cr
    t         & \epsilon_R(\vec{k})    },
\label{hamiltMatrix}
\end{equation}
where $\epsilon_{L(R)}(\vec{k})$ are single-well eigenenergies,
\begin{equation}
\epsilon_{L(R)}(\vec{k}) = \frac{\hbar^2}{2m^*}
\left(k_x-k_0\right)^2 + \frac{\hbar^2k_y^2}{2m^*}\,,
\label{eigene}
\end{equation}
and $k_0 = \pm|e|Bd/2\hbar$.  The hopping integral $t$ is given by a
matrix element of the well potential between lowest eigenstates of the
left and right wells.  The diagonalization of the matrix
(\ref{hamiltMatrix}) yields the bonding and antibonding eigenstates.

We employ a minimal model that accounts for the inter-well
polarization effect in the Hartree approximation. Transport
coefficients are obtained using the random phase approximation and
Kubo formalism of linear response theory.  In our geometrical
arrangement the ``parallel'' conductivity $\sigma_{\|}$ relates the
current component $j_y$ to the external electric field component $E_y$
in a standard way where $j_y=\sigma_{\|} E_y$.  The structure of
$\sigma_{\perp}$ is more complicated as it involves the
electron-electron interaction contribution. The basic scheme of our
calculation is outlined below; for more details see
Ref.\ ~\cite{Kol}. The ``perpendicular'' conductivity is composed of
$\sigma_{\perp}^{(1)}$, which is the response to the applied field
component $E_x$, and $\sigma_{\perp}^{(2)}$  corresponding to the
``Hall'' electric field $E_z$, i.e., $j_x=
\sigma_{\perp}^{(1)}E_x+\sigma_{\perp}^{(2)}E_z$.  The ``Hall'' field
appears as a consequence of the polarization of the sample and is
related to the inter-well ``Hall'' potential $U_H$ as $E_z =
-U_H/d$. It is calculated in a self-consistent way from the
non-equilibrium excess charge $\delta Q$ as a solution to the Poisson
equation. In our model, $\delta Q$ and the potential $U_H$ are related
by a particularly simple formula corresponding to the parallel plate
capacitor,
\begin{equation}
 \delta Q  = \varepsilon \frac{U_H}{d},
\label{charge}
\end{equation} 
where $\varepsilon $ denotes the dielectric constant of the barrier.
The excess charge $\delta Q$ is obtained as a sum of components
$\delta Q^{(1)}$ and $\delta Q^{(2)}$ calculated from the Kubo formula
as a response to the fields $E_x$ and $E_z$, respectively. Writing
\begin{equation}
 \delta Q^{(1)} = \varepsilon^{(1)} E_x ,\,\,\,\,
 \delta Q^{(2)} = \varepsilon^{(2)} E_z ,
\end{equation}
where $\varepsilon^{(1)}$ and $\varepsilon^{(2)}$ are the generalized
dielectric functions,  $E_z$ is the solution to the self-consistent
equation (\ref{charge}) and takes the form
\begin{equation}
E_z  = \frac{\varepsilon^{(1)}}{\varepsilon^{(2)}
+\varepsilon}\, E_x .
\label{ez} 
\end{equation}
This formula relates the self-consistent Hartree field of electrons to
the $x$\,~component of the applied external electric field
$\vec{E}$. Using Eq.\,~(\ref{ez}), the perpendicular conductivity can
be written in a form
\begin{equation}
\sigma_{\perp} = \sigma_{\perp}^{(1)}
  + \frac{\varepsilon^{(1)}}{\varepsilon^{(2)}
+\varepsilon}\, \sigma_{\perp}^{(2)} .
\label{delta}
\end{equation}
In this equation the first term $\sigma_{\perp}^{(1)}$ corresponds to
the conventional solution of the Boltzmann equation while the second
term represents the novel ``Hall'' correction to the conductivity.
\section{Short-range scatterers in the Born approximation}
We have performed numerical calculations of transport coefficients for
the realistic parameters of a bilayer 2D system.  Since we have only a
limited knowledge of the nature of scatterers and their distribution
in the sample, we assume that electrons are scattered by impurities
distributed randomly in both left and right wells. To be more
specific, the scattering on an individual impurity is considered to be
intra-well (diagonal in the layer index) and isotropic in the
$\vec{k}$ space.  The concentration of impurities is assumed very low
and the weak scattering on an individual impurity is treated in the
non-self-consistent version of the Born approximation. The strength of
the scattering on an individual impurity and the impurity
concentrations are taken as adjustable parameters.

In the course of calculation we introduced the resolvents in
expressions for the dielectric functions and the conductivity
components.  The finite quasiparticle lifetime and the transport
relaxation time are related to the imaginary part of the self-energy
and to the vertex corrections to velocity components, respectively.
The quantities result from replacing the resolvents and their
products, characteristic for a given configuration of scatterers, by
expressions averaged over all possible configurations.

To proceed further, we adopt additional simplifications justified by
the very low scattering rate in high-mobility samples.  Only terms of
the lowest order in the scattering rate are kept in all
expressions. While $\sigma_{\|}$, $\sigma_{\perp}^{(1)}$ and
$\varepsilon^{(1)}$ would diverge in samples without impurities, the
leading terms, $\sigma_{\perp}^{(2)}$ and $\varepsilon^{(2)}$, of the
expansions obtained as a response to $U_H$, are finite in this limit.
Note that, in spite of finite $\sigma_{\perp}^{(2)}$, both terms of
$\sigma_{\perp}$ are inversely proportional to the scattering rate, as
the second term is proportional to $\varepsilon^{(1)}$ according to
Eq.\,~(\ref{delta}).
 
The vertex corrections vanish for the velocity component $v_y$ as
expected for randomly distributed short-range scatterers.  This is not
true for $v_x$.  Due to the anisotropy of the scattering induced by the
in-plane magnetic field, the vertex corrections have to be taken into
account.
\section{Results and discussion}
The in-plane magnetic field changes qualitatively the topology of
Fermi contours.  At zero field, the Fermi contours are two concentric
circles. The in-plane field shifts the centers of Fermi circles in
opposite directions in $k$-space and gradually changes the contour
shapes.  At the critical field $B_{c,1}$, the Fermi line of an
antibonding subband vanishes and the transition from a two-component
to a one-component system occurs. The Fermi contour of a bonding
subband splits at the higher critical field $B_{c,2}$ into two parts
and the left and right electron layers are decoupled.
\begin{figure}[htb]
\includegraphics[angle=0, width=7.5cm]{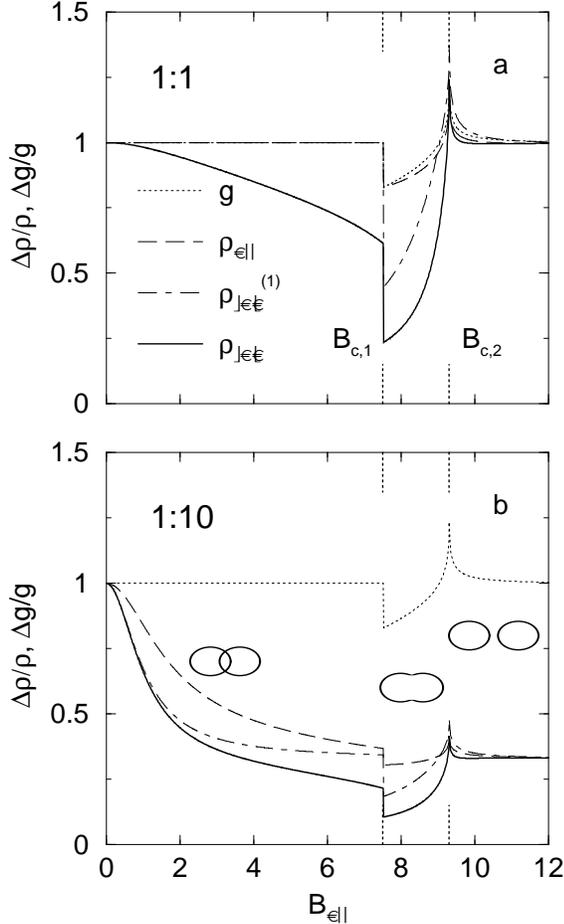}
\caption{Calculated field dependence of the density of states $g$ and
the magnetoresistance curves $\rho_{\|}$ and $\rho_{\perp}$. In
$\rho_{\perp}^{(1)}$ the ``Hall'' contribution is neglected. (a) The
concentration of impurities is the same in both wells. (b) The ratio
of concentrations in two wells is 1:10, i.e. the scattering occurs
mostly in one of the wells. The schematic shape of the Fermi contours
below, between and above the critical fields $B_{c,1}$ and $B_{c,2}$
is also shown}
\label{fig2}
\end{figure}
The calculated magnetoresistance components $\rho_{\|}$ and
$\rho_{\perp}$, are presented in Fig.~\ref{fig2}, together with the
field dependence of the density of states $g$. The van Hove
singularities at  critical fields $B_{c,1}$ and $B_{c,2}$ seen on
$g$ appear also on $\rho_{\|}$ and $\rho_{\perp}$. Two different
distribution of the scattering centers are considered, a symmetric
case (a) where the concentration of impurities is identical in both
wells, and an asymmetric case (b) where  the prevailing amount of
impurities is in one of the wells and the ratio of their concentrations
is 1:10. Three regions of the magnetic fields should be discussed
separately: below $B_{c,1}$, between $B_{c,1}$ and $B_{c,2}$ and above
$B_{c,2}$.

In Fig.~\ref{fig2}a, the curves are strikingly  similar  below $B_{c,1}$,
except of $\rho_{\perp}$. 
The components $\rho_{\|}$ and $\rho_{\perp}^{(1)}$,
which can be obtained by solving the Boltzmann equation, exactly
coincide with $g$ for our simplified model. It is the Hall
contribution which qualitatively changes the field dependence of
$\rho_{\perp}$ to negative magnetoresistance in this region.  Between
$B_{c,1}$ and $B_{c,2}$ only the bonding subband is occupied and the
difference between $\rho_{\|}$ and $\rho_{\perp}$ originates from the
Fermi contour anisotropy and the related anisotropy of the effective
mass and the relaxation time. The anisotropy of the transport
relaxation time (the vertex correction to $v_x$) plays a role also
above $B_{c,2}$.

Fig.~\ref{fig2}b shows the magnetoresistance curves calculated for the
asymmetric distribution of scattering centers. The dramatic reduction
of all magnetoresistance components with increasing magnetic field is
related to the in-plane field suppression of the coupling between
wells. The tunneling between wells is reduced with increasing magnetic
field and the conductivity of the well with lower concentration of
impurities prevails. In this case the ``Hall'' correction does not
change the field dependence qualitatively but enhances the
difference between $\rho_{\|}$ and $\rho_{\perp}$ substantially.

To conclude, we have shown that the charge redistribution within a
double-layer system gives rise to the Hall-like contribution to the
perpendicular magnetoresistance component $\rho_{\perp}$ which is so
strong that it cannot be omitted in the analysis of bilayer 2D system
magnetotransport. 
\section*{Acknowledgements}
This work has been supported by the Grant Agency of the Czech
Republic under Grant No. 202/01/0754.

\end{document}